\documentclass[preprint, onecolumn, 1p, number]{elsarticle}
\usepackage{natbib}
\usepackage{graphicx}
\usepackage{epstopdf}
\usepackage{listings}
\usepackage{amsmath}
\usepackage{placeins}
\journal{Journal of Computational Physics}
\begin{document}
\setlength{\unitlength}{1.0mm}
\begin{frontmatter}
\title{Parallel exact diagonalization solver for quantum-electron models}
\author{S. N. Iskakov}
\ead{iskakoff@ngs.ru}
\author{V. V. Mazurenko}
\ead{mvv@dpt.ustu.ru}

\address{Ural Federal University, Mira St. 19, 620002 Yekaterinburg, Russia}

\begin{abstract}
We present a parallel computation scheme based on the Arnoldi algorithm for exact diagonalization of quantum-electron models.
It contains a selective data transferring method and distributed storage format for efficient computing of the
matrix-vector product on distributed computing systems. The performed numerical experiments demonstrated good
performance and scalability of our eigenvalue solver. The developed technique has been applied to investigate the
electronic properties of Sr$_{2}$RuO$_4$ at experimental temperatures. The role of the spin-flip term in the electronic
Hamiltonian was analyzed.
\end{abstract}
\begin{keyword}
DMFT \sep sparse matrices \sep matrix-vector product \sep distributed computing
\end{keyword}
\end{frontmatter}

\section{Introduction}
Dynamical mean-field theory (DMFT) has been successfully used for studying the electronic properties of a variety of
strongly correlated materials in which the strength of the electron-electron interaction energy is comparable to or
larger than the kinetic one \citep{RevModPhys.68.13}.  Since these characteristics lead to a wealth of physical
phenomena such as metal insulator transitions, exotic magnetic structures, and unconventional superconducting phases
\citep{RevModPhys.70.1039}, the DMFT method attracts a great deal of attention. The main underlying idea of this
approach is the mapping of the lattice problem onto an effective impurity problem which in turn is solved numerically
exactly. For that one can use different numerical techniques such as  exact diagonalization (ED), numerical
renormalization group (NRG), quantum Monte Carlo method with the Hirsch-Fye algorithm (QMC-HF), or other schemes
\citep{RevModPhys.68.13}.

Many interesting and promising results were obtained by using QMC methods. For instance, a realistic five-band
modeling of the strongly correlated materials is mainly performed by using QMC Hirsch-Fye method. However, there is a
number of technical limitations. (i) The simulation temperature of 250 K - 1000 K is much higher than
experimental one of 10 K - 100 K. (ii) The Coulomb interaction matrix contains density-density terms only. (iii) The
sign problem prevents scientists from calculating the physical properties at sufficiently low temperatures
\cite{PhysRevB.80.155132}. (iv) The QMC approaches require a maximum entropy method for analytical continuation of the
resulting Green's function to real-energy axis.

Such problems can be solved  by using the ED techniques. The full diagonalization method is limited to three-band
systems because of the extremely rapid increase of the Hilbert space with the number of the electronic states $N_s$.
For instance, full diagonalization of the Hamiltonian with two impurity orbitals, each coupled to three bath levels
($N_s=8$), requires diagonalization of matrices with dimension up to $4900$ which roughly represents the time and
storage limit of what is computationally meaningful.

The diagonalization problem  can be simplified by exploiting the extreme sparseness of the Hamiltonian
matrices and focusing on the limited number of the states with the lowest energies. It can be done by using
Lanczos \cite{capone:245116} or Arnoldi \cite{perroni:045125} approaches. As a result the Hamiltonian corresponding to
$N_s=12$ can be treated at about the same computational cost as the full diagonalization for $N_s=8$. This improvement
allows the application of ED/DMFT to realistic three-band  and five-band systems \citep{perroni:045125}. However, there
is memory limitation of the Hamiltonain matrix storage to apply ED/DMFT to more complicate physical systems. To date the
biggest matrix that can be diagonalized by this approach on a modern workstation is about 6435$^2$ $\times$ 6435$^2$
which corresponds to $N_s$ =15 \cite{PhysRevB.81.054513}.

In this paper we present a sparse matrix-vector product (SpMV) parallelization strategy for effective solving ED/DMFT
equations on distributed computing systems. For these purposes we suggest two improvements of the original ED/DMFT
method \cite{RevModPhys.68.13}: a distributed storage format for sparse large matrix and a selective data transferring.
The former allows the application of the ED/DMFT method for system with the number of electronic states more than
$N_s=15$. The selective data transferring method substantially reduces amount of data transmitted between the
processors of a distributed computing system, which improves SpMV performance.

The paper organized as follows. In section 2, we presents the basic steps of the DMFT theory. In section 3,
the developed eigenvalue solver and its parallel performance are given. We also give approbation of our computational
scheme for a one-band Hubbard model on square lattice. For purpose of illustration in section 4 we will use a
three-band model for Sr$_2$RuO$_4$ compound.

\section{Theory}
One way to describe many-particle phenomena such as high-temperature superconductivity, heavy fermions and others is to solve the
Hubbard Hamiltonian \cite{Hubbard1963}
\begin{equation}
\label{eqn:math:1.1}
H = - \sum_{i, j, \sigma} t_{ij} c_{i\sigma}^{+}c_{j\sigma}  + U \sum_i n_{i\uparrow}n_{i\downarrow}
\end{equation}
where $c_{i\sigma}^{+}\left(c_{i\sigma}\right)$ is creation (annihilation) operators of fermion, $t_{ij}$ is a hopping
integral, $U$ is the on-site local Coulomb interaction, $n_i$ is the particle number operator. It is a very complicated
problem due to the exponential growth of the Hamiltonian and, hence, the direct solution of the Hubbard Hamiltonian is
only possible for small model clusters \cite{1105808}.

In case of the real systems one can use the Dynamical Mean-Field Theory \cite{RevModPhys.68.13} in which the lattice
problem is reduced to a single-site effective problem. Thus, such a mean-field consideration leads to freezing spatial
correlations. Instead of the Hubbard Hamiltonian we solve the Anderson impurity model which is given by
\begin{equation}
\label{eqn:math:1.2}
H_{AIM} = \sum_{k,\sigma}V_{kd\sigma}\left[d_{\sigma}^{+} c_{k\sigma} + c_{k\sigma}^{+} d_{\sigma}\right]
+\sum^{N_k}_{k,\sigma}\varepsilon_{k\sigma} c_{k\sigma}^{+} c_{k\sigma} + \varepsilon_{d} (n_{\uparrow}+n_{\downarrow})
+ U n_{\uparrow} n_{\downarrow},
\end{equation}
where $N_k=N_{s}-1$ is a number of the bath states, $d^{+}_\sigma$($d_\sigma$) and $c_{k\sigma}^{+}$ ($c_{k\sigma}$) are
creation (annihilation) operators for fermions with spin $\sigma$ associated with the impurity site and with the state
$k$ of the effective bath, respectively, $\varepsilon_k$ and $\varepsilon_d$ are the energy of bath and impurity,
$V_{kd}$ is a hybridisation integral of impurity and bath states. The main task of the ED/DMFT is to diagonalize the
impurity Hamiltonian $H_{AIM}$ to compute the impurity Green function which can be written as
\begin{eqnarray}
\label{eqn:math:1.3}
&G\left(i\omega_n\right) = \frac{1}{Z}\sum_{\nu\mu}\frac{\|\langle\mu\|c_{\sigma}^{+}\|\nu\rangle\|^2}
{E_{\nu} - E{\mu} - i\omega_n}\left[e^{-\beta E_{\nu}} + e^{-\beta E_{\mu}}\right], 
\end{eqnarray}
where $\mid \nu \rangle$ ($\mid \mu \rangle$) and $E_{\nu}$ ($E_{\mu}$) are eigenvectors and eigenvalues of $H_{AIM}$
and $Z = \sum_{\mu} \exp(-\beta E_\mu)$ is the partition function. The exponential factor in (\ref{eqn:math:1.3})
indicates that at large enough $\beta$ only a small number of the lowest eigenstates needs to be calculated
\cite{capone:245116}. In this paper, we develop a high performance parallel technique for solving the eigenvalue
problem of the Anderson impurity model Hamiltonian matrix (\ref{eqn:math:1.2}) on distributed computing systems.

\section{Distributed eigenvalue solver}
A few iterative numerical algorithms such as a power method, a Lanczos method, an implicitly restarted Arnoldi method
\cite{maschhoff96portable}, the conjugate gradient (CG) method \cite{Stiefel1952} have been proposed to solve the
eigenvalue problem for large sparse matrices. Most of them are based on iterative multiplication of the Hamiltonian
matrix $H$ by the trial vector $\nu$. Thus they are ideally suited for parallelization.

An efficient implementation of the Lanczos method for symmetric multiprocessing computers with large shared memory 
was proposed in Ref.\cite{1353039} for solving the Heisenberg model. The authors have demonstrated a perfect
scalability of their computational scheme. It was possible since they used a SMP machine with eight cores which
were sitting on the same board. As we will show below it is not the case for a distributed architecture.

Recently, the ED results for a five-band DMFT problem were reported by Liebsch \cite{PhysRevB.81.054513}. Judging by a
short description of the technical details  they have also used a SMP machine with large shared memory and 32
processors on the board. The aim of our work is to modify the exact diagonalization method for solving DMFT equations
on distributed computing systems.

\subsection{Distributed CSR format and selective data transferring}
Since the Hamiltonian matrix $H$ is extremely sparse the performance of the $H\times\nu$ operation strongly depends on
the storage format and SpMV algorithm we use. 
The memory subsystem and, more specifically, the memory bandwidth is identified as the main performance bottleneck of
the SpMV routine \cite{1366244,331600}. This performance bottleneck is due to the fact that SpMV performs $O(nnz)$
operations where $nnz$ is a number of the nonzero elements of the sparse matrix, which means that most of the data are
accessed in a streaming manner and there is little temporal locality. To avoid indirect memory access some sparse
matrix storage format is used. One of the most widely used storage format for sparse matrices is
Compressed Sparse Row (CSR) format \cite{Mellor-crummey03optimizingsparse, Im:CSD-00-1104} which stores all the
non-zero elements of the Hamiltonian in contiguous memory locations ($H_{values}$ array) and uses two additional arrays
for indexing information: $row\_ind$ contains the start of each row within the non-zero elements array and $col\_ind$
contains the column number associated with each non-zero element. \begin{figure}[tbh]
\centering
$H = \left(\begin{matrix}
h_{00} & 0 & 0 & 0 & h_{04} & 0 \\
0 & h_{11} & 0 & 0 & 0 & h_{15} \\
0 & 0 & h_{22} & 0 & 0 & 0 \\
0 & 0 & 0 & h_{33} & 0 & 0 \\
h_{40} & 0 & 0 & 0 & h_{44} & 0 \\
0 & h_{51} & 0 & 0 & 0 & h_{55}
\end{matrix}\right)$

{\Huge$\Downarrow$}

$row\_ind = \left(\begin{smallmatrix} 0 & 2 & 4 & 5 & 6 & 8 & 10 \end{smallmatrix}\right)$

$col\_ind = \left(\begin{smallmatrix} 0 & 4 & 1 & 5 & 2 & 3 & 0 & 4 & 1 & 5 \end{smallmatrix}\right)$

$H_{value} = \left(\begin{smallmatrix}
h_{00} & h_{04} & h_{11} & h_{15} & h_{22} & h_{33} & h_{40} & h_{44} & h_{51} & h_{55}
\end{smallmatrix}\right)$

\caption{Example of CSR storage format. The original matrix H is compressed into three arrays $row\_ind$, $col\_ind$
and $H_{value}$.}
\label{fig:fig1}
\end{figure}
The size of the $H_{values}$ and $col\_ind$ arrays are equal to the number of the non-zero elements ($nnz$), while the
size of the $row\_ind$ array is equal to the number of rows ($nrows$) plus one. An example of the CSR format for a
sparse $6 \times 6$ matrix is presented in Figure \ref{fig:fig1}. In order to organize the matrix-vector multiplication
operation one can use the following listing (Listing \ref{CommonSRCSpMV}).

\begin{lstlisting}[caption=Sparse Matrix-Vector Multiplication using CSR format, label=CommonSRCSpMV]
for (i=0; i<Hdim; i++)
  for (j=row_ind[i] ; j<row_ind[i+1]; j++)
    y[i] += H[j] * v[col_ind[j]];
\end{lstlisting}

In the framework of the standard SpMV procedure with the standard CSR format all three arrays and vector should be
stored on the same board. It is impossible when the size of the Hamiltonian is larger than the memory of a single
machine. Therefore the first goal of our work is to design method for arrays distribution over CPU mesh to overcome the
problem of large matrix storage.

We use the standard CSR storage format as a starting point. The compressed Hamiltonian as well as the accompanied
arrays are distributed over processors grid. It reduces required memory space of a single CPU (table \ref{Table1}). The
Hamiltonian matrix is stored distributively all the time. For this purpose we have developed the library written in C++
for initialization and working with different elements of distributed arrays.

\begin{figure}[tbh]
\centering
\includegraphics[width=0.8\textwidth,angle=0]{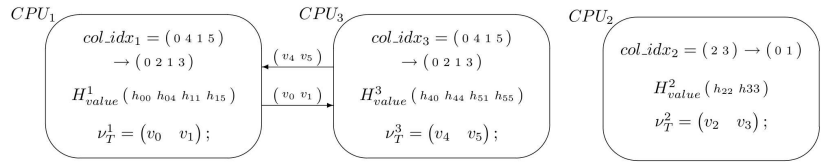}
\caption{ Example of Distributed CSR storage format running on 3 CPU. The arrows denote the communications between CPUs.}
\label{fig:fig2}
\end{figure}

The second goal of our work is to organize effective and optimal communications between processors to reduce amount of
transmitted data. To solve this problem we have developed communication procedures for operating elements of the matrix
stored distributively. The subroutines can be divided into two classes. The first one is used for exchanging data
between CPUs based on one-sided communication functions of MPI. For optimal performance and memory usage we used
MPI\_Get function, because in case of exchanging large amount of data MPI\_Put subroutine needs significantly more
memory. The second class of the subroutines is directed to determine groups of interacting CPUs. In a sense the main
advantage of the distributed CRS format we proposed is that each CPU is to communicate with only a few CPUs to receive
the different parts of the Arnoldi vector during simulation. For example, while solving eigenvalue problem for the
matrix with the size of $165,636,900$ only about $30$ of a $256$ CPUs are needed for each CPU to receive required parts
of the vector.

The figure \ref{fig:fig2} shows an example of distributed CSR format for $6 \times 6$ matrix in case of three
processors. One can see that the original compressed matrix (Fig.\ref{fig:fig1}) and the array $col\_idx$ is
distributed into 3 arrays. Since the matrix-vector multiplication each CPU needs only a small part of the whole
vector then the index array can be considerably reduced. Thus the approach we proposed naturally leads to reduction of
single CPU memory requirement for the trial vector storage. CPU 1 and CPU 3 are to transmit the different parts of the
vector to each other.

\subsection{Performance}

\begin{figure}[tbh]
\centering
\includegraphics[width=0.8\textwidth,angle=0]{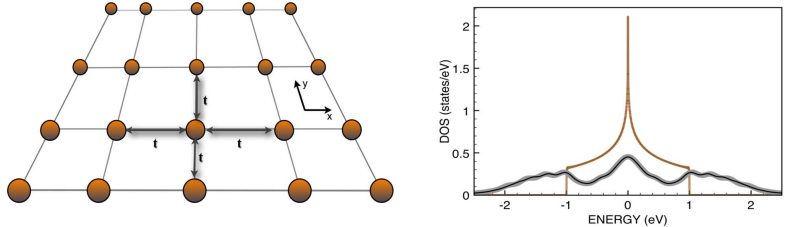}
\caption{(left) Schematic representation of the square lattice with the nearest neighbors couplings. (right) Spectral
functions obtained from ED/DMFT calculations with $N_s$ =16 (gray bold line). The
dashed brown line corresponds to the non-interacting density of states.}
\label{fig:fig3}
\end{figure}

To test the performance of the developed technique we have performed the ED/DMFT calculations for a one-band Hubbard
model on a square lattice with the on-site Coulomb interaction $U$ = 2 eV and the nearest hopping integral $t$ = 0.5
eV (Fig.\ref{fig:fig3}). The calculations were carried out for $N_s$=16. The resulting Green function is presented in
Fig. \ref{fig:fig3}. One can see that there is the famous three peaks structure (quasiparticle peak, lower and upper
Hubbard bands) that agrees with results of the previous theoretical investigations \cite{RevModPhys.68.13}.

Table \ref{Table1} shows the performance of the modified ED/DMFT calculation scheme on $N_{p}$ = 32, 64, 128, 256, and
512 processors. These performance measurements were made as follows. The total elapsed time and time of communications
between processors were measured by MPI\_Wtime function. The required memory was measured by the tools installed on a
particular parallel cluster system.

\begin{figure}[tbh]
\centering
\includegraphics[width=0.8\textwidth,angle=0]{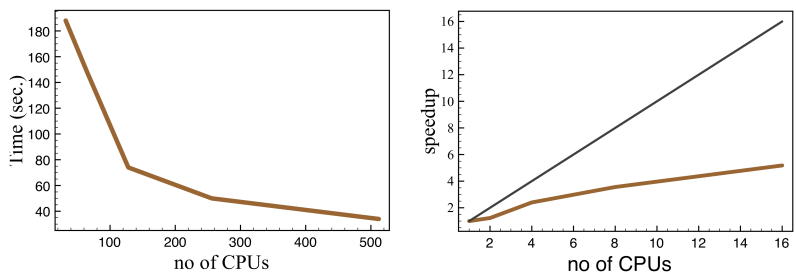}
\caption{Scaling of the total computation time for 1 ED/DMFT iteration with number of used CPUs. The number of CPU in
the speedup dependence (right) is normalized by 32.}
\label{fig:fig4}
\end{figure}

Fig.\ref{fig:fig4} shows the scaling of the computation time for 1 ED/DMFT iteration with a maximum vector length of
11,778,624. One can see that the speedup of our calculation scheme is far from the ideal scaling. The main reason for
that is time required for communication between different CPUs. Another operation of our computational scheme which
takes much time is the construction of the Hamiltonian matrix  (Table \ref{Table1}).

We can also compare the speedup of our algorithm in case of the different sizes of the Hamiltonian matrix. For $N_s$=16
the ratio $T_{256}/T_{512}$ = 1.58 is larger and closer to the ideal one than $T_{32} / T_{64}$ =1.26 in case of
$N_{s}$=14. Based on the obtained performance data we can estimate the number of processors we need for $N_{s}$ = 18
and $N_{s}$ =20. For instance, in case of $N_{s}$ = 18, for the largest sector one needs to store about the 82 billions
of non-zero elements. It requires about 1.5 Tb of memory and can be solved by using 2000 cores cluster.
Investigation of the systems with $N_{s}$ = 20 leads to operating about 1194 billions of the non-zero elements and it is
impossible to solve this problem using our scheme.

It is interesting to compare the performance of the developed technique with that reported in Ref.
\cite{PhysRevB.80.165126}. For $N_{s}$ =12 the authors of the paper have estimated that the time of the one iteration
with 300 lowest eigenstates was less than 30 min if 16 processors were employed in parallel. In our case the similar
calculations take us about 45-50 min. Such a time difference can be explained by taking into account that the author
used a workstation with shared memory and processors on the same board.

\begin{table}[h!]
\caption{Performance of the developed eigenvalue solver. $N_p$ is a number of processors and $nnz_{max}$ is a maximum
number of the  non-zero elements per row, $M_{exch}$ and $M_{data}$ are amount of transferred data per processor and
memory requirement per processor (in MB). $T_{total}$, $T_{comm}$ and $T_{setup}$ are the total computational,
communication time and time for generating Hamiltonian matrix (in sec.), respectively.}
\begin{center}
\begin{tabular}[t]{*{9}{|c}|}
\hline
$N_s$& H$_{dim}$ & $nnz_{max}$ & $N_p$ & M$_{exch}$ & M$_{data}$ & $T_{setup}$ & $T_{comm}$ & $T_{total}$\\
\hline
14&11,778,624&13&32&23&300&17  &66    & 188\\
14&11,778,624&13&64&22&150&10  &71    & 149 \\
14&11,778,624&13&128&15&75&3.2    &47    & 74 \\
14&11,778,624&13&256&7&40&1.7    &35    &  50\\
14&11,778,624&13&512&4&20&1.0    &22    & 34\\
16&165,636,900&15&256&120 &500  &52  &330 & 602 \\
16&165,636,900&15&512&54&270 &25 &200  & 382 \\
\hline
\end{tabular}
\end{center}
\label{Table1}
\end{table}

It is important to discuss the possible ways to improve our computation scheme. Firstly, The results of the previous
work \cite{Schnack20084512} have demonstrated that it would be better not to store the Heisenberg matrix, but to evaluate
the matrix elements whenever needed. Secondly, since we divide the full Hamiltonian matrix into sub-matrices of equal
number of the elements there is an irregular distribution of the non-zero elements. It results in unequal partition of
workloads among processors. This load-imbalance problem can be solved by partitioning  the Hamiltonian matrix in a
computational space where the optimal number of the non-zero elements is defined to minimize the load-balance and
communication costs.

\section{Numerical results}
Using the developed exact diagonalization technique we have investigated the electronic structure of Sr$_{2}$RuO$_4$
compound. This system is of interest due to a number of unconventional physical properties. For instance,
Sr$_{2}$RuO$_4$ demonstrates an unconventional superconductivity at temperatures below 1 K. Moreover,  photoemission
and optical experiments have shown substantial mass renormalization effects. In the previous theoretical investigations
the electronic structure of Sr$_{2}$RuO$_4$ was studied within LDA+DMFT method using the QMC-HF as the impurity solver.
For instance, Liebsch and Lichtenstein \cite{PhysRevLett.84.1591} have performed multiband DMFT calculations to explain
the discrepancy between photoemission and de Haas-van Alphen data. It was found that there is a charge transfer from
$3d_{xz, yz}$ states to $3d_{xy}$ states. The similar QMC-HF method was used by the authors of the paper
\cite{PhysRevB.75.035122} to describe some features of photoemission spectra.

In previous investigations the lowest simulation temperature was 15 meV, which is larger those at which experiments
were performed.  The imaginary time QMC data were analytically continued by the maximum entropy method. Since the
QMC-HF/DMFT method operates only with density-density interaction terms the spin-flip term in the impurity Hamiltonian
was not taken into account. At the same time the effect of such an interaction on electron spectra was mentioned in a
number of works \cite{PhysRevLett.84.1591}.

To take into account the spin-flip effects in our ED/DMFT calculation we have diagonalized the following impurity
Hamiltonian:
\begin{eqnarray}
H = \sum_{m \sigma} (\epsilon_{m} -\mu) n_{m \sigma} + \sum_{k \sigma} \epsilon_{k} n_{k \sigma} + \sum_{m k \sigma} V_{m k}
[d^{+}_{m \sigma} c_{k \sigma} + H.c. ] + \sum_{m} U n_{m \uparrow} n_{m \downarrow} \nonumber \\ + \sum_{m < m' \sigma \sigma'} (U' - J \delta_{\sigma \sigma'} ) n_{m \sigma} n_{m' \sigma'} 
- \sum_{m \ne m'} J' [d^{+}_{m \uparrow} d_{m \downarrow} d^{+}_{m' \downarrow} d_{m' \uparrow} + d^{+}_{m \uparrow} d^{+}_{m \downarrow}
d_{m' \uparrow} d_{m' \downarrow} ].
\end{eqnarray}
Here $d^{(+)}_{m \sigma}$ are annihilation (creation) operators for impurity electrons on orbital $m$ with spin
$\sigma$, $U (U')$ is the intra-orbital (inter-orbital) impurity Coulomb energy, J is an intra-atomic exchange integral
and $J'$ is the spin-flip interaction. The account of the latter interaction is possible in the framework of the exact
diagonalization and recently developed CT-QMC calculation scheme. In according with Ref.\cite{PhysRevLett.84.1591}
these parameters were chosen to be $U=U'=1.2$ eV, $J=0.2$ eV and $J'=0.1$ eV. The simulation temperature was taken to
be T =200 K that is close to the experimental values used in Ref.\cite{PhysRevLett.84.1591}. Thus for a total of about
20 excited states the Green’s functions are calculated via the Lanczos procedure. Of course, this number increases at
higher temperatures, and if Boltzmann factors smaller than $10^{−5}$ are included for higher precision.

\begin{figure}[tbh]
\centering
\includegraphics[width=0.8\textwidth,angle=0]{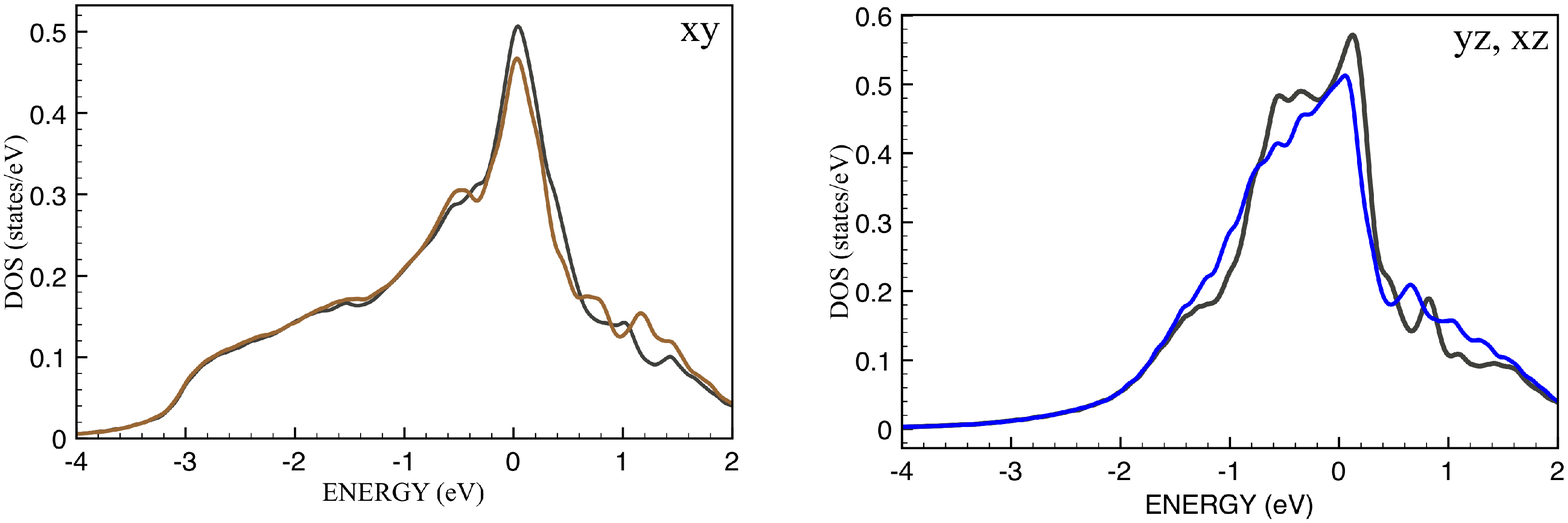}
\caption{Densities of states obtained from DMFT calculations  for T = 200 K with (brown and blue lines) and without
(gray lines) the spin-flip term.}
\label{fig:fig5}
\end{figure}

The resulting Green functions obtained for $N_s$=15 is presented in Fig.\ref{fig:fig5}. The spectral functions without
the spin-flip term agree well with those presented in Ref. \cite{PhysRevLett.84.1591}. One can see that the account of
the spin-flip term leads to an additional renormalization of the spectrum near the Fermi level.
It means that other non-diagonal elements of the full U-matrix can play an important role for correct description 
of the experimentally observed spectra. The approach we used here is also well suited for analyzing the ground and few
excited states. Such an analysis can help us to study the experimentally observed superconducting phase.

\section{Conclusions}
To conclude, we developed a distributed storage format and a selective data transferring procedure for distributed
sparse matrix-vector multiplication. They were implemented in ED solver for dynamic mean field theory problem and
showed good speedup and scalability. The obtained numerical results look quite promising. The developed approach can be
used for solving Heisenberg model on computer clusters with distributed memory.

\section{ACKNOWLEDGMENTS}
We would like to thank A.O. Shorikov for his technical assistance with test calculations.  
This work is supported by the scientific program ``Development of
scientific potential of universities'' N 2.1.1/779, by the
scientific program of the Russian Federal Agency of Science and
Innovation N 02.740.11.0217, the grant program of President of
Russian Federation MK-1162.2009.2., RFFI 10-02-00546.
The calculations have been performed on the cluster of University Center for Parallel Computing of Ural Federal University and 
the Brutus cluster at ETH Zurich.

\bibliographystyle{elsarticle-num-names}
\bibliography{iskakov_mazurenko}

\end{document}